\documentclass[aps,prc,twocolumn,groupedaddress,showpacs,showkeys]{revtex4}
\usepackage{graphicx}
\begin{document} 
\title{Statistical Model Predictions for Particle Ratios at $\sqrt{s_{NN}}$ = 5.5 TeV}
\author{J.~Cleymans${}^{1}$, I.~Kraus${}^{2}$, H. Oeschler${}^{2}$, K.~Redlich${}^{3}$ and S.~Wheaton${}^{1,2}$}
\affiliation{${}^{1}$UCT-CERN Research Centre and Department  of  Physics,\\ University of Cape Town, Rondebosch 7701, South Africa\\
		${}^{2}$Institut f\"ur Kernphysik, Darmstadt University of Technology, D-64289 Darmstadt, Germany\\
		${}^{3}$Institute of Theoretical Physics, University of Wroc\l aw, Pl-45204 Wroc\l aw, Poland}
\date{\today}
\begin{abstract} 
Particle production in central Pb-Pb collisions at LHC is discussed in the context of the Statistical Model. Predictions of various particle ratios are presented with the corresponding choice of model parameters made according to the systematics extracted from heavy-ion collisions at lower energies. The sensitivity of several ratios on the temperature and the baryon chemical potential is studied in detail, and some of them, which are particularly appropriate to determine the chemical freeze-out point experimentally, are indicated. We show that the $\rm \bar{p} / \rm p$ ratio is most suitable to determine the baryon chemical potential while the $\Omega^-$/K$^-$ and  $\Omega^-$/$\pi^-$ ratios are best  to determine the temperature at chemical freeze-out.
\end{abstract}

\maketitle
\section{Introduction} 
Particle production observed in heavy-ion collisions allows a systematic study of the thermal properties of the final state. In a wide energy range, from the SIS up to RHIC, the  yields of produced particles have been  shown to be consistent with the assumption that hadrons originate from a thermal source with a given temperature and a given baryon density. These  properties have been  quantified by comparing the measured particle ratios with Statistical Model calculations. By using only  two thermal parameters, a successful description of particle ratios measured in heavy-ion  collisions over a wide range of center of mass energies could be made~\cite{review}. The extracted chemical freeze-out parameters, the temperature $T$ and the baryon chemical potential $\mu_B$ can be characterized by a constant average energy per hadron $\langle E \rangle / \langle N \rangle$ of approximately 1 GeV~\cite{freeze-out}. The extrapolation of this freeze-out curve towards vanishing $\mu_B$ is bounded by the critical phase transition temperature as calculated in Lattice Gauge Theory~\cite{lattice}.

In view of  the success of the Statistical Model,  we discuss in this paper the expectations for particle ratios  in central heavy-ion collisions at the LHC energy. The calculations are performed at the chemical decoupling point extrapolated from the  freeze-out systematics. The particle ratios are discussed with respect to their sensitivity to variations in the thermal parameters. Based on this analysis, we suggest observables  which are best suited to extract experimentally the freeze-out conditions in heavy-ion collisions at the LHC energy. 

At LHC energies, due to the large number of produced particles, a novel type of data analysis will be possible. In particular, the thermal origin and chemical decoupling conditions in heavy-ion collisions can be  studied for  various event classes and even within a  single event. Consequently, the freeze-out conditions can be extracted event-wise, leading to distributions of the temperature and baryon chemical potential determined  event-wise. Instead of a full analysis within the Statistical Model, the proposed particle ratios can be utilized to extract these values.

Furthermore, at these energies most particles might be produced via hard collisions which could lead to particle ratios substantially different from those of  the Statistical Model. Deviations from these predictions might lead to new insight into the hadronisation mechanism.
\section{Statistical Model} 
The Statistical Model and its application to heavy-ion collisions has been summarized recently~\cite{review}. In the present calculations we  apply the Statistical Model with completely equilibrated flavor production. Due to  the  expected large multiplicity  of strange particles in central Pb-Pb collisions at LHC energies, we use the grand-canonical formulation of strangeness conservation. Thus, strangeness,  baryon number and  electric charge are determined by corresponding chemical potentials. It should be noted that for very peripheral heavy-ion collisions, the suppression of the strange-particle phase space due to canonical effects might occur even at LHC energies.
\\
The charge chemical potential is constrained by the initial isospin asymmetry of the Pb nuclei, whereas the strange chemical potential $\mu_S$, depending on both $T$ and $\mu_B$, is determined by strangeness neutrality.  Thus, any particle ratio is uniquely determined by only two parameters, $T$ and $\mu_B$ at chemical freeze-out.

In order to estimate particle ratios in Pb-Pb collisions at LHC, one needs to determine the expected range of thermal parameters. For this purpose  we apply the parameterized $\sqrt{s_{NN}}$-dependence  of $\mu_B$ as it was recently extracted from the Statistical Model analysis of hadron multiplicities in heavy-ion collisions \cite{parametrization}. The baryon chemical potential $\mu_B$, extrapolated to the LHC energy, is, according to this parameterization,
\begin{equation}
\mu_B (\sqrt{s_{NN}} ={\rm 5.5~TeV)~=~1 ~MeV.}
\end{equation}
Similar estimates of $\mu_B$ at the LHC energy have been obtained recently in \cite{cleymans_extra,pbm_lecture,pbm_paper,becattini}.
\\
Considering the variation  of the chemical freeze-out temperature with collision energy in heavy-ion collisions, evidently $T$ is an increasing function of $\sqrt{s_{NN}}$. However, from SPS to RHIC the change of the chemical freeze-out temperature is only moderate. Within the statistical and systematic errors  the temperature $T\approx 170$ MeV is consistent with the value derived at the top  SPS and RHIC energies~\cite{review,T_SPS1,T_SPS2,T_RHIC1,T_RHIC2}. Lattice QCD results on the critical temperature, extrapolated to the chiral limit and at vanishing baryon density, concur within statistical error with this value~\cite{lattice}. Thus, we use this value as an estimate of the chemical freeze-out temperature at LHC. 
\\
However, to account for possible uncertainties in  the extrapolations of the chemical freeze-out parameters to the LHC energy, we consider a possible shift of $T$ by $\pm$5~MeV and a variation of $\mu_B$ between 0 and 5 MeV. 

From the sensitivity of different particle ratios on freeze-out parameters we deduce  the reliability of the model predictions and the ability to determine the parameters experimentally. To identify the relevant particle ratios that can be used to constrain experimentally the  chemical decoupling conditions in heavy-ion collisions at the LHC energy, we consider a broader range of parameters. The  ratios are studied for $150 {\rm ~MeV}<T<180 {\rm ~MeV}$  and  for $0<\mu_B< 30{\rm~MeV}$.
\section{Particle ratios at LHC} 
%
%
%
\begin{table}
\caption{\label{Table} Particle ratios in central Pb-Pb collisions at freeze-out conditions expected at the
					LHC: $T$ = (170$\pm$5) MeV  and $\mu_B$~=~${\rm 1^{+4}_{-1}}$ MeV. 
					The given errors correspond to the variation in the thermal parameters.
					Additional, systematic uncertainies in the ratios of the right column arise from
					unknown decay modes. They are smaller than 1\% in general, but reach 3\% in 
					the $\Xi^- / \Lambda$ ratio and 7\% in the  p$ / \pi^-$ and the $\Lambda / $p  
					ratios.}
\begin{ruledtabular}
\begin{tabular}{cc|cc}
\multicolumn{2}{c|}{ $\bar{h} / h$ Ratio } & \multicolumn{2}{c}{ mixed Ratio } \\ 
\hline
 $ \pi^+ / \pi^- ~ $ & ~ $ 0.9998^{+ 0.0002}_{- 0.0010}$ ~ & ~ $ \rm K^+ / \pi^+ ~ $ & ~ $ 0.180^{+ 0.001}_{- 0.001} $~  \\
 $ \rm K^+ / \rm K^- ~ $ & ~ $ 1.002^{+ 0.008}_{- 0.002}$ ~ & ~ $ \rm K^- / \pi^-~ $ & ~ $ 0.179^{+ 0.001}_{- 0.001} $~  \\
 $ \rm \bar{p} / \rm p	~ $ & ~ $ 0.989^{+ 0.011}_{- 0.045}$ ~ & ~ $ \rm p / \pi^-	~ $ & ~ $ 0.091^{+ 0.009}_{- 0.007} $~  \\
 $ \bar{\Lambda} / {\Lambda}~ $ & ~ $ 0.992^{+ 0.009}_{- 0.036}$ ~ &  ~ $ \Lambda / \rm p ~ $ & ~ $ 0.473^{+ 0.004}_{- 0.006} $~  \\
 $ \bar{\Xi}^+ / {\Xi}^-	~ $ & ~ $ 0.994^{+ 0.006}_{- 0.026}$ ~ &  ~ $ \Xi^- / \Lambda	~ $ & ~ $ 0.160^{+ 0.002}_{- 0.003} $~  \\
 $ \bar{\Omega}^+ / {\Omega}^-	~ $& ~ $ 0.997^{+ 0.003}_{- 0.015}$ ~ & ~ $ \Omega^- / \Xi^-~ $ & ~ $ 0.186^{+ 0.008}_{- 0.009} $~  \\
%
\end{tabular}
\end{ruledtabular}
\end{table}
With the values of the thermal parameters expected at LHC and discussed in the previous Section we  calculate a set of particle ratios using the {\sc Thermus} package~\cite{thermus}. The predicted particle ratios for Pb-Pb collisions at LHC are summarized  in Table~\ref{Table} and~\ref{Table2}. The errors arise from the variation of the temperature and of the baryon chemical potential, as indicated. Additional systematic uncertainties, not included in the Tables, are due to incomplete knowledge of  decay modes of resonances into stable particles. We estimate that  due to the uncertainty in branching ratios, the  $\rm p / \pi^-$ and the $\Lambda / \rm p$ ratios might vary by up to 7\%, whereas for the  $\Xi^- / \Lambda$ ratio a change of 3\% is possible. The other ratios are less sensitive to the resonance properties and affected by less than 1\%.  Within the considered range of thermal parameters, the antiparticle/particle ratios are insensitive to any variation in contributions from resonances, due to the decay chains being equivalent  for particles and antiparticles.
\begin{table}
\caption{\label{Table2} Left column: Resonance/stable particle ratios in central Pb-Pb collisions at freeze-out conditions 
					expected at the
					LHC: $T$ = (170$\pm$5) MeV  and $\mu_B$~=~${\rm 1^{+4}_{-1}}$ MeV. 
					The given errors correspond to the variation in the thermal parameters.
					Additional, systematic uncertainies arise from unknown decay modes. 
					They are smaller than 1\% in general, but reach 2\% in 
					the $\phi / \rm K^-$ ratio and 3.5\% in the  $\rho^0(770) / \pi^-$  
					ratio. 
					Right column: Resonance width, taken from \cite{pdg}.}
\begin{ruledtabular}
\begin{tabular}{cc|cc}
\multicolumn{2}{c|}{ Ratio } & \multicolumn{2}{c}{ Width (MeV) } \\ 
\hline
 $ \phi / \rm K^- ~ $ 				& ~ $ 0.138^{+ 0.004}_{- 0.004}$ ~ & $\phi$ 				& ~ $ 4.26 \pm 0.05 $~  \\
 $ \Lambda(1520) / \Lambda ~ $ 		& ~ $ 0.090^{+ 0.003}_{- 0.003}$ ~ & $\Lambda(1520)$ 		& ~ $ 15.6 \pm 1.0 ~$~  \\
 $ \rm K^\ast(892)^0 / \rm K^- ~ $ 	& ~ $ 0.323^{+ 0.010}_{- 0.009}$ ~ & $\rm K^\ast(892)^0$ 	& ~ $ 50.7 \pm 0.6 ~$~  \\
 $ \rho^0(770) / \pi^- ~ $ 			& ~ $ 0.127^{+ 0.001}_{- 0.002}$ ~ & $\rho^0(770)$ 		& ~ $ 150.3 \pm  1.6 ~ ~$~  \\
%
\end{tabular}
\end{ruledtabular}
\end{table}

One of the expected features of particle production at LHC is that there should be a rather negligible difference between the yield of particles and their antiparticles. This is a direct consequence of the low net-baryon density expected in the collision fireball at LHC at chemical decoupling. With the range of the chemical potential $\rm 0<\mu_B<5  ~{\rm MeV}$ used in the actual calculations, the ratios of particle to antiparticle yields, called $\bar{h} / h$ hereafter, are indeed seen in Table~\ref{Table} to be close to unity. Figure~\ref{fig1} illustrates the dependence of $\bar{h}/h$ ratios on $T$ (left) and on $\mu_B$ (right). The calculations at $\mu_B \approx \rm 27~MeV$ serve as cross checks with experimental data. At RHIC, at $\sqrt{s_{NN}} \rm = 200~GeV$, several  particle ratios were measured in Au-Au reactions by the BRAHMS~\cite{brahms}, PHENIX~\cite{phenix}, PHOBOS~\cite{phobos} and STAR~\cite{star} collaborations. Good agreement is observed for all ratios under study. As an example the $\rm \bar{p}/p$ ratio is indicated in Figure~\ref{fig1}.

On the other hand, the ratios of particles with different masses and different quantum numbers are predominantly controlled by the freeze-out temperature and the particle masses (see Sec.~\ref{sens} below). The values of such yield ratios vary strongly with the particle mass difference, as seen in the right panel of Table \ref{Table}. 
\\
Some of the ratios are expected  to change only slightly at LHC from their values measured at the SPS and RHIC. This can be seen from Fig.~\ref{fig2}, showing little sensitivity of these ratios with $T$ and $\mu_B$. A particular example are the K$/ \pi$ ratios which, within errors, are consistent with those measured at lower energies. Large deviations from SPS and RHIC measurements are seen for particle ratios built with baryons and antibaryons. This is a direct consequence of a substantial decrease in baryon stopping at LHC, which, in the Statistical Model, results in a decreasing baryon chemical potential. 

Particle densities in central Pb-Pb collisions at $\sqrt{s_{NN}} ={\rm 5.5~TeV}$ have been recently calculated within relativistic hydrodynamics~\cite{eskola}. Their results on ratios of pions, kaons and protons are comparable to our findings given in Table~\ref{Table}. However, the predictions of this hydrodynamic model depend strongly on the value of the  decoupling temperature. A similar sensitivity on the decoupling temperature is found in both models.

Resonance yields were suggested as a sensitive probe of the fireball expansion dynamics~\cite{rafelski}. 
The resonances might be sensitive to collective effects and interactions with the  surrounding medium. Such interactions can result in resonance broadening or mass shifts~\cite{brown,rapp}. Assuming that resonances are produced according to a thermal distribution, their initial multiplicity can be calculated. Deviations from the thermal yields can be attributed to processes which take place in a later stage of the evolution after chemical freeze-out. In this late stage, the expansion dynamics might be driven mostly by elastic and pseudo-elastic hadron scattering. Also the reconstruction of short-lived resonances which decay in this stage will fail if at least one daughter particle undergoes  an interaction with the medium. On the other hand, pseudo-elastic processes might regenerate resonances~\cite{markert} and consequently, the yield of resonances at thermal freeze-out can differ from the one calculated in the Statistical Model at chemical freeze-out. This has been observed in measurements of resonance yields in heavy-ion collisions at RHIC~\cite{markert}. The observed yields of  short-lived resonances like $\rho$ or $K^*$ were found to differ from the model predictions even by a factor of two~\cite{review}.  Benchmarks are essential to possibly evaluate modifications in resonance yields. Therefore, resonance/stable particle ratios, as expected at chemical freeze-out, are calculated in the Statistical Model without including dynamical effects in the final state. They are displayed in Table \ref{Table2} together with the width of the resonances. Yields of shorter lived  resonances, i.~e. those with a large width, are expected to be affected more strongly~\cite{markert}.
\section{\label{sens}Sensitivity to freeze-out parameters} 
In this section we discuss the dependence of different particle ratios on the values of the thermal parameters. The main objective of this study is to identify observables  that could serve as sensitive experimental probes for the chemical freeze-out conditions at the LHC energy. In this context we consider the antiparticle to particle ratios as well as the ratios of strange and non-strange particles of different masses and quantum numbers.
\subsection{Antiparticle to Particle Ratios} 
%
%
\begin{figure}
\includegraphics[width=9.5cm]{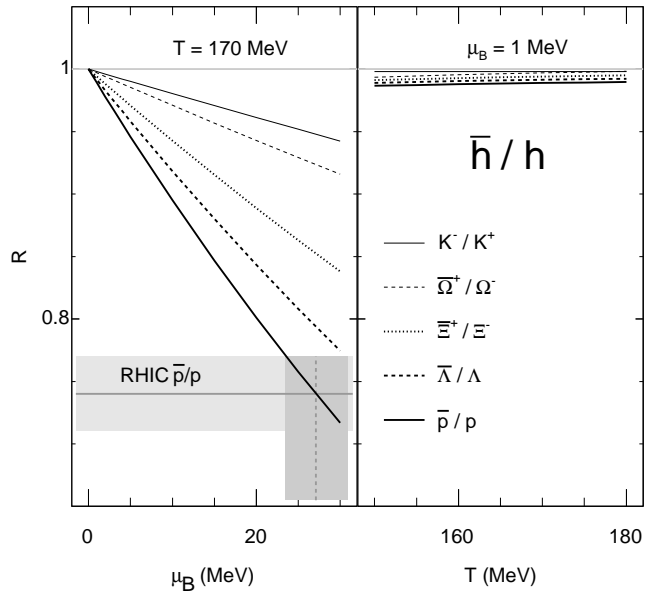}
\\ ~ \vspace{-11mm} \\ 
\caption{\label{fig1}Antiparticle/particle ratios $R$ as a function of $\mu_B$ for $T$ = 170~MeV (left) and as a function of $T$ for $\mu_B$ = 1~MeV (right). The horizontal line at 1 is meant to guide the eye. The $\rm \bar{p}/p$ ratio (averaged over the data of the 4 RHIC experiments at $\sqrt{s_{NN}}$~=~200~GeV) is displayed (gray horizontal line) together with its statistical error (gray band). As illustrated, $\mu_B \approx \rm 27~MeV$ (dashed line) can be read off the Figure directly within the given accuracy (vertical gray band).} 
\end{figure} 
Figure~\ref{fig1} shows the dependence of different antiparticle/particle ratios, $\bar h/h$, on $\mu_B$ (left) and $T$ (right). For vanishing baryon chemical potential, the density of particles is identical to that of antiparticles, thus $\bar h/h=1$. For finite and increasing baryon chemical potential, the $\bar h/h$ ratios are decreasing functions of $\mu_B$. Such  properties of antiparticle/particle ratios are qualitatively well understood in the Statistical Model. The mass terms in the single particle and antiparticle partition functions are identical. Consequently, the leading dependence of the $\bar h/h$ ratios on $\mu_B$  is determined by,
\begin{equation}
\bar{h}/h \propto \exp[-2(B \mu_B + S \mu_S)/T], 
\label{equ2}
\end{equation}
where $B$ and $S$ are the  baryon  and strangeness quantum numbers of the particle, respectively. In the  expression above, feed-down contributions from resonance decays  are ignored but they have been included in the model calculations. However, due to equivalent contributions of resonance decays to particles and antiparticles, Eq.~(\ref{equ2}) provides a good description of $\bar h/h$ ratios at the LHC energy. In the antibaryon/baryon ratios, the $\mu_B$-term  dominates over the $\mu_S$-term because strangeness conservation implies that the baryon chemical potential is, in a strangeness neutral system and at a fixed temperature, always larger than the strangeness chemical potential. This, together with the opposite sign of the quantum numbers $B$ and $S$ of  strange baryons, results in a weaker sensitivity on $\mu_B$ of $\bar{h}/h$ ratios for hadrons with increasing strangeness content. Consequently, as seen in Fig.~\ref{fig1}, at a fixed value of $\mu_B$, the antibaryon/baryon ratios are increasing with strangeness content of the hadrons. 

The $\rm K^-/ K^+$ ratio shows qualitatively a similar dependence on the baryon chemical potential as the antibaryon/baryon ratios. This is due to the strangeness content of the kaons and the fact that $\mu_S$ is an increasing function of $\mu_B$.  The $\rm K^-/K^+$ ratio has no explicit dependence on the baryon chemical potential and is therefore always larger than the antibaryon/baryon ratios at the same $\mu_B$ and $T$.

For small values of $\mu_B$, as expected in heavy-ion collisions at the LHC energy, the $\bar h/h$ ratios  are only  weakly dependent on  temperature,  as seen in Fig.~\ref{fig1} (right). Thus, these ratios do not  constrain the chemical freeze-out temperature in heavy-ion collisions. On the other hand, the  high sensitivity of $\bar h/h$ on $\mu_B$ makes such ratios ideal observables to  quantify $\mu_B$ experimentally. As seen in Fig. 1, this is particularly  the case for the  $\rm \bar{p}/p$ ratio which exhibits the strongest dependence on the baryon chemical potential.
\subsection{Mixed Particle Ratios} 
%
%
\begin{figure}
\includegraphics[width=9.5cm]{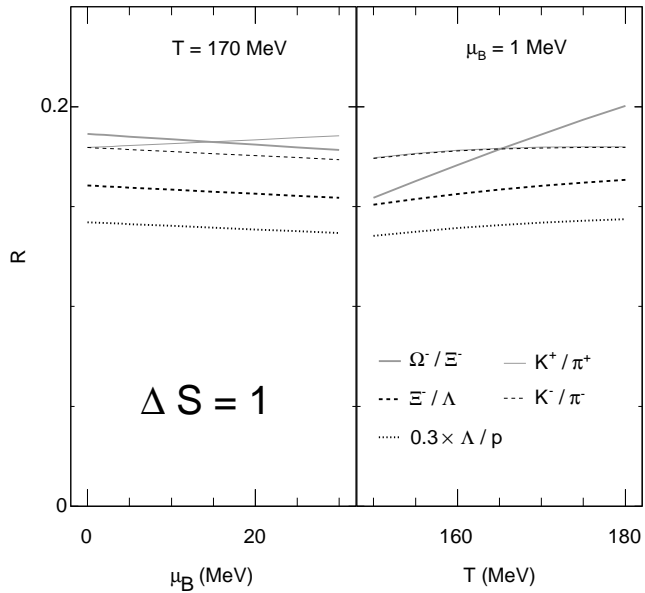}
\\ ~ \vspace{-11mm} \\ 
\caption{\label{fig2}
Ratios $R$ of particles with unequal strangeness content as a function of $\mu_B$ for $T$ = 170 MeV (left) and as a function of $T$ for $\mu_B$ = 1 MeV (right).} 
\end{figure} 
From the preceding Section one can conclude that ratios of particles with equal masses are not good thermometers of the medium. To extract the chemical freeze-out temperature in heavy-ion collisions, one should consider ratios that are composed of hadrons having different  masses. The dependencies of such ratios on $\mu_B$ and $T$ are illustrated   in Fig.~\ref{fig2}. Within the range of the freeze-out parameters considered in this figure, there is a rather weak sensitivity  of   the ratios on $T$ and $\mu_B$. In the  case of the $\Omega^- / \Xi^-$ ratio  this can be seen analytically from the expression,
\begin{equation}
\frac{\Omega^-}{\Xi^-} \propto \Bigl(\frac{m_{\Omega^-}}{m_{\Xi^-}} \Bigr) ^{3/2} {\rm exp} \Bigl[- \frac{m_{\Omega^-} - m_{\Xi^-}}{T} \Bigr]  ~ {\rm exp} \Bigl[ - \frac{ \mu_S} {T} \Bigl] ,
\label{equ3}
\end{equation}
where, in this simplified formula, feeding from resonance decays is ignored, with  $m_{\Omega^-}$ and $m_{\Xi^-}$ denoting the corresponding particle masses.
\\
There is no explicit dependence on the baryon chemical potential in all ratios considered in Fig.~\ref{fig2}, as also seen in Eq.(\ref{equ3}). Changes in $\mu_B$ affect the ratios only indirectly through the $\mu_B$-dependence of the strangeness chemical potential. The values of $\mu_S$ are nearly proportional to $\mu_B$ in the range studied here. However, for all values of $\mu_B$ considered in Fig. 2, it is always  smaller than 8~MeV. Consequently, the dependence on the baryon chemical potential is comparatively weak.
\\
The temperature dependence of the $\Omega^- / \Xi^-$ ratio is dominated by the mass term, since the difference ($(m_{\Omega^-} - m_{\Xi^-})\approx 350$~MeV) is much larger than the strangeness chemical potential ($\mu_S<1$ MeV for $\mu_B\approx 1$ MeV and for all considered values of $T$). However,  to correctly quantify the temperature dependence of  the ratios of particles with different masses, one  needs to  include contributions from resonance decays.

%
%
\begin{figure}
\includegraphics[width=8.5cm]{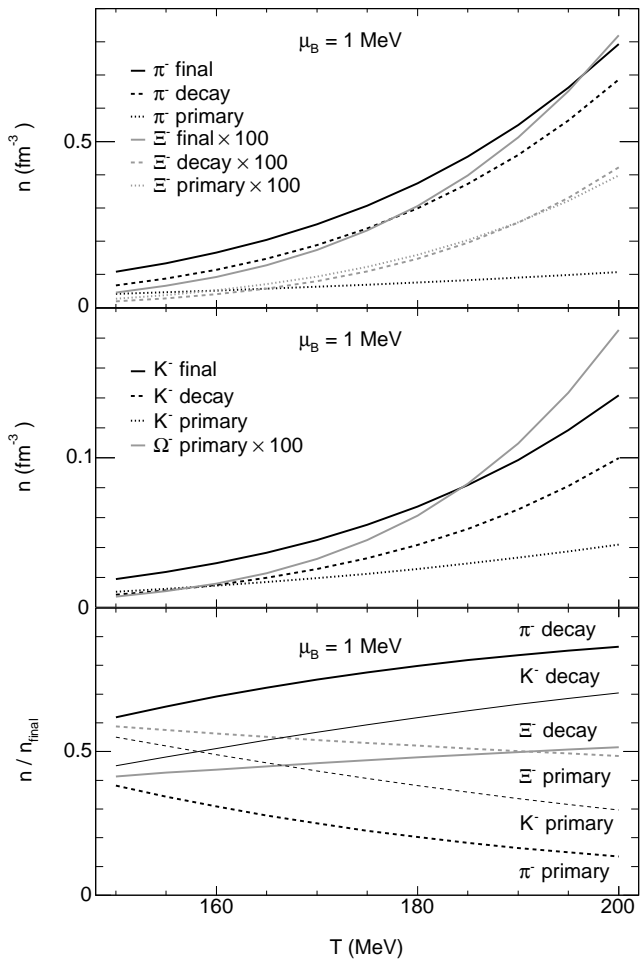}
\caption{\label{fig3}Upper and middle panel: Particle densities of $\pi^-$, $\Xi^-$, K$^-$ and $\Omega^-$ hadrons as a function of $T$ for $\mu_B$ = 1 MeV. Different sources are indicated. The thermally produced, primary hadrons together with those from decays sum up to the finally observed densities. The primary and final density of $\Omega^-$ hyperons is identical, (i.e. there is no contribution from resonances).\\
Lower panel: Fractional amount of $\pi^-$, K$^-$ and $\Xi^-$ hadrons from primary production and from decays of the final densities, shown as a function of $T$ for $\mu_B$ = 1 MeV.} 
\end{figure}
Figure~\ref{fig3} (upper and middle panel) displays particle densities from different sources. Particles called {\sl primary} are directly produced from a thermal system with  given temperature and chemical potentials. The second contribution, called {\sl decay} in  Fig.~\ref{fig3}, originates from short-living resonances which feed the yield of stable hadrons. The sum of both contributions results in the {\sl final} particle density.
\\
There is no feeding to $\Omega^-$ hyperons from heavier resonances. In contrast, only about 50\% of the $\Xi^-$ hyperons are primary, while the other half originates from resonance decays. Since both contributions exhibit a similar temperature dependence, as demonstrated in Fig.~\ref{fig3}, the sensitivity of the $\Omega^- / \Xi^-$ ratio on $T$ can still be estimated from~Eq. (\ref{equ3}). The $\Omega^- / \Xi^-$ ratio increases by about 1\% per 1~MeV temperature increase.
\\
For light hadrons like kaons and pions, the feed-down contributions exceed the thermally produced yield. As seen in Fig.~\ref{fig3} (lower panel) for the thermal conditions at LHC, approximately 75\%  of the pions are expected to be produced by resonance decays. Furthermore, the contributions from resonance decays feature a temperature dependence deviating from that of primarily created hadrons, as  seen in  Fig.~\ref{fig3}; the former exhibit a steeper increase. Therefore, the temperature-driven increase in the heavy/light hadron ratios is progressively diluted when  hadrons with lower masses are considered in the denominator. In the  case of the K/$\pi$ ratios, the contribution to the total  pion yield from decays compensates fully for an increase in the kaon yield with increasing temperature. Consequently, the K/$\pi$ ratios are almost independent of temperature.
\\
This observation allows a stringent test of the model. If the experimental values for the K/$\pi$ ratios deviate significantly from the predictions, they can hardly be reconciled with the Statistical Model. This would be an indication for a non-equilibrated system at chemical freeze-out.

%
%
\begin{figure}
\includegraphics[width=9.0cm]{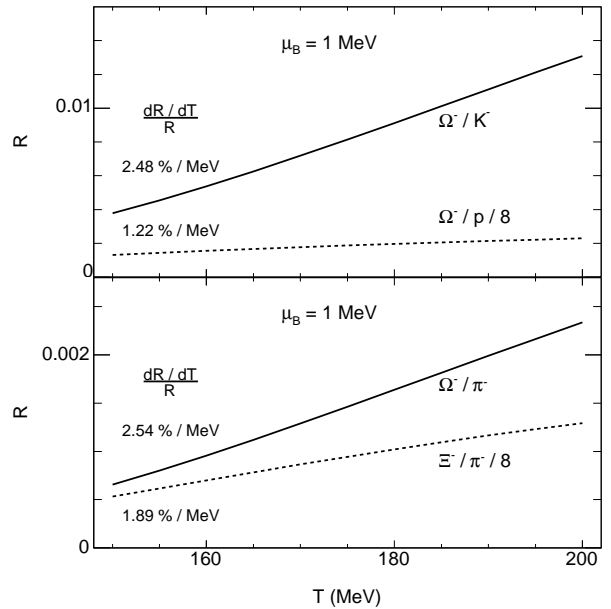}
\caption{\label{fig4}
Particle ratios $R$ involving hyperons as a function of  $T$ for $\mu_B$ = 1 MeV.} 
\end{figure} 
From the discussions above and in the previous Section, it becomes clear that neither antiparticle/particle ratios nor ratios composed of hadrons with small mass differences are well suited to deduce the chemical freeze-out temperature. However, ratios with hyperons, in particular the hyperon/pion ratios, are  excellent observables to extract the chemical decoupling temperature in heavy-ion collisions. Figure~\ref{fig4} shows the temperature  dependence of some particle ratios containing  hyperons. The Statistical Model predictions of these ratios are shown up to $T~=~200$ MeV. 
\\
In general,  an upper limit for the chemical freeze-out temperature at LHC is given by the value of the QCD critical temperature  at vanishing chemical potential. A recent lattice QCD finding~\cite{katz} indicates that in (2+1)--flavor QCD the critical temperature can be even as large as 195~MeV. This value is significantly larger than previously expected \cite{lattice}. Figure~\ref{fig4} illustrates that the particle ratios are sensitive to temperature variations in the range that includes the above value of the critical temperature. However, at temperatures larger than 190~MeV, the Statistical Model results start to suffer from an imprecise knowledge of high-mass resonances and their decay properties.
\\
The sensitivity of different particle ratios on temperature  is quantified in Fig.~\ref{fig4}  by  the variation of the ratio $R$ per 1 MeV change in $T$ relative to the value of $R$ calculated at $T$ = 170 MeV. The results given in this figure show that the temperature dependence is strongest  for particle ratios with a large mass difference. This statement  is also valid for ratios composed of antiparticles.
\\
The largest  (2.5\%/MeV) sensitivity on temperature is exhibited by  the $\Omega^- / \pi^-$ ratio. However,  the  variation is significantly smaller than that expected from only thermally produced particles. This is because at LHC only about 25\% of all pions are expected to be directly produced, while the remaining fraction originates from baryonic and, predominantly, from mesonic resonances~\cite{spencer}. The feed-down contribution to pions increases with temperature and reduces the sensitivity.
\\
In view  of  the   large decay contributions to the  pion yield, the $\Omega^-/$K$^-$ ratio might be a better thermometer of the medium, since the feeding to kaons is noticeably smaller. Figure~\ref{fig4} demonstrates that the two ratios, $\Omega^- / \pi^-$ and $\Omega^-/$K$^-$, are similarly sensitive to temperature variations. The combination of the larger kaon mass and smaller feeding compared to pions results in comparable  temperature dependencies. 
\\
The $\Xi^- / \pi^-$  ratio has  a  similar mass difference as the $\Omega^-/$K$^-$ ratio. However, the very large  feeding of heavy resonances to pions destroys the expected similarity in the temperature dependence of these ratios. The sensitivity of the hyperon to proton ratio on temperature is rather weak due to the small mass difference of constituent particles.

From Fig. \ref{fig4} it is obvious that the $\Omega^- / \pi^-$ and $\Omega^-/$K$^-$  ratios  are  best suited  to determine the chemical freeze-out temperature. Experimentally, the latter might be more easily accessible and more precisely known, since  the contributions from weakly decaying particles, which have to be obtained with high accuracy, are larger for $\pi^-$ than for K$^-$ mesons.
\section{Conclusion and Summary} 
Predictions of the Statistical Model for different particle ratios for Pb-Pb collisions at the LHC energy are presented. The sensitivities of various ratios with respect to the temperature and the baryon chemical potential as well as the contribution of resonances  are discussed and analyzed. We have shown that  the $\rm \bar{p} / \rm p$ ratio is the best suited observable to extract the value of the baryon chemical potential at chemical freeze-out.  The $\Omega^- / \pi^-$ and the $\Omega^-$/K$^-$ ratio are proposed as thermometers to extract experimentally the chemical freeze-out temperature in central Pb-Pb collisions at LHC.

Hard processes could  dominate the particle production at LHC energies. From these interactions, very different particle ratios than those of a thermal source might be expected to be created. Deviations from the predictions given here will help to understand the dynamics and, possibly, the time scale of the hadronisation procedure. Furthermore, in events selected e.g.~by high-energy jets, the resulting particle ratios might be different, indicating perhaps a higher freeze-out temperature or perhaps deviating completely from the thermal picture. Such selection criteria might make it possible to differentiate between  early and late freeze-out. On the other hand, the frequent interactions between the constituents close to the phase boundary~\cite{wetterich} could lead to a complete reshuffling and a statistical distribution might be observed at the end, independent of the preceding history.
\end{document}